\documentstyle[12pt]{article} 
 \newlength{\extraspace}
 \newlength{\extraspaces}
\newcommand{\be}{\begin{equation}
\addtolength{\abovedisplayskip}{\extraspaces}
\addtolength{\belowdisplayskip}{\extraspaces}
\addtolength{\abovedisplayshortskip}{\extraspace}
\addtolength{\belowdisplayshortskip}{\extraspace}}
\newcommand{\ee}{\end{equation}}

\newcommand{\bear}{\begin{eqnarray}
\addtolength{\abovedisplayskip}{\extraspaces}
\addtolength{\belowdisplayskip}{\extraspaces}
\addtolength{\abovedisplayshortskip}{\extraspace}
\addtolength{\belowdisplayshortskip}{\extraspace}}
\newcommand{\ear}{\end{eqnarray}}
\newcommand{\lsi}{\raise0.3ex\hbox{$<$\kern-0.75em\raise-1.1ex\hbox{$\sim$}}}
\newcommand{\gsi}{\raise0.3ex\hbox{$>$\kern-0.75em\raise-1.1ex\hbox{$\sim$}}}

\begin{document}
\begin{flushright} HD--THEP--01--35
\end{flushright}


{\thispagestyle{empty}
\begin{center}\vspace*{1.0cm}
{\Large Are Galaxies Cosmon Lumps?} \\

\vspace*{1.0cm} {\large Christof Wetterich} \\
\vspace*{0.5cm}{\normalsize {Institut f\"ur Theoretische Physik, \\
Universit\"at Heidelberg, \\ Philosophenweg 16, \\ D--69120 Heidelberg,
Germany.}} \\

\end{center}
\begin{abstract}
The scalar ``cosmon'' field mediating quintessence
influences the dynamics of extended objects in the universe.
We discuss cosmon lumps -- spherically symmetric solutions for
the scalar field coupled to gravity. The two integration constants
can be associated to the mass and the rotational velocity in a halo-like
region with constant rotation curve. The presence of the scalar field
also changes the singularity of the black hole solution.
We ask if galaxies could be associated with cosmon lumps.
\end{abstract}

\noindent
Quintessence is an interesting and perhaps the most natural candidate
for the dark energy in the universe \cite{1}, \cite{2}. It is mediated
by a scalar field -- the cosmon -- which is driven by a decaying
potential to infinity. The dark energy of the universe is associated
to the potential and kinetic energy of the coherent motion of
the cosmon field and therefore homogeneously distributed in the universe
\cite{1}, \cite{2}. For an appropriate form of the effective cosmon potential
and kinetic term the equation of state of quintessence may have a
negative pressure today. This could explain many of the recently observed
surprising features of the universe \cite{3}.
Quintessence influences the age of the universe \cite{1},\cite{2},
the effective equation of state of the energy momentum tensor
\cite{1},\cite{2}, the formation of structure \cite{2}, \cite{4},
\cite{4a},
\cite{5} and the detailed fluctuation pattern of the cosmic
microwave background \cite{2},\cite{4},\cite{6},\cite{7}. A negative
pressure in the present epoch
reconciles a large amount of dark energy  with
structure formation and explains the cosmological acceleration
as observed from distant supernovae \cite{8}.

All this concerns the role of the homogeneous field which can be
interpreted as the space average of the cosmon field. The cosmon
is essentially massless on scales smaller than the horizon \cite{1}.
Therefore, there are also important questions about the role it
could play on the scale of galaxies and clusters. Could the incoherent cosmon  
fluctuations on small scales play the role of dark matter \cite{CDM}
and drive the formation of structure in the universe?
Which role plays the cosmon field in a galaxy or a cluster?
If the cosmon has a weak coupling to baryonic matter or to some
unknown particle constituting the dark matter \cite{1}, the
local cosmon fluctuation has necessarily a nonzero value in any
concentration of matter. In this case a source term drives the local field
away from the cosmological value \cite{9}. Even without such a coupling
the presence of a local cosmon field in extended objects seems
quite plausible. Due to its long-range character this field would
not vanish outside a matter concentration. The cosmon coupling
to gravity modifies then the gravitational solutions for
``empty space'' around local objects.

We speculate in this note that the cosmon could even dominate
the energy momentum tensor of extended objects. If so, one may
in a first approximation neglect the matter component and investigate
the coupled system of cosmon and gravity.
More generally, the ``matter'' (e.g. baryons) of
an extended object may be concentrated in an inner
``bulk region''. Outside of the bulk the solution of the
cosmon-gravity field equations becomes relevant for all
models with quintessence.
We discuss\footnote{For solutions of the Einstein-Klein-Gordon
field equations in the context of Bose stars see ref. \cite{BS}.}
spherically symmetric solutions with intriguing properties
ressembling galactic halos with a constant rotation curve. The outer
region of galaxies could be associated to such a cosmon
lump, whereas the inner region  needs
the inclusion of matter. In a more conservative scenario dark
matter would be an essential component of the halo. Still,
the cosmon field may play a quantitatively important role.
In particular, its
presence modifies the region around the central singularity.

For the standard isotropic metric $ds^2=-B(r)dt^2+A(r)
dr^2+r^2(d\theta^2+\sin^2
\theta d\varphi^2)$ the gravitational field equations in presence of
a scalar field $\varphi(r)$ are given by
\bear\label{1}
R_{00}&=&\frac{B''}{2A}-\frac{B'}{4A}\left(\frac{A'}{A}+\frac{B'}{B}\right
)+\frac{B'}{rA}=-\frac{BV}{2M^2}\nonumber\\
R_{rr}&=&-\frac{B''}{2B}+\frac{B'}{4B}\left(\frac{A'}{A}+\frac{B'}{B}\right
)+\frac{A'}{rA}=\frac{AV+{\varphi'}^{2}}{2M^2}\nonumber\\
R_{\theta\theta}&=&1+\frac{r}{2A}\left(\frac{A'}{A}-\frac{B'}{B}\right
)-\frac{1}{A}=\frac{r^2V}{2M^2}\ear
Here primes denote derivatives with respect to $r$ and $M^2=M^2_p/16\pi
=(16\pi G)^{-1}$. For simplicity of the discussion we assume here that  
quintessence is characterized by
a standard kinetic term and an exponential potential
\be\label{2}
V=M^4\exp(-\alpha\frac{\varphi}{M})\ee
such that the scalar field equation reads
\be\varphi''+\left(\frac{2}{r}-\frac{A'}{2A}+\frac{B'}{2B}\right)
\varphi'=A\frac{\partial V}{\partial\varphi}=-
\frac{\alpha}{M}AV\ee
We will present a class of solutions for which the
precise form of the potential term is
actually not important for the qualitative behavior.
Other potentials for quintessence may therefore
be used as well, provided that they share the rough features characteristic
for quintessence.

We concentrate first on a free scalar field with $V=0$. Appropriate
linear combinations of the field equations yield two coupled first-order
differential equations for $A$ and
\be\label{3a}
w=\frac{r}{M}\varphi'\ee
namely
\be\label{4}
rw'=-Aw\ ,\quad rA'=-A(A-1-\frac{w^2}{4})\ee
The general solution depends on two integration constants $R_s$
and $\gamma$
\be\label{5}
w=\gamma\frac{R_s}{\rho}\ ,\quad A=1+\frac{R_s}{\rho}-\frac{\gamma^2}
{4}\left(\frac{R_s}{\rho}\right)^2\ee
and the new radial coordinate $\rho$ is related to $r$ by
\be\label{6}
\frac{\partial\ \ln\rho}{\partial\ \ln r}=A\ee
It obeys
\be\label{7}
r=(\rho-\rho_H)^{\frac{1}{2}-\delta}(\rho+\rho_H+R_s)^{\frac{1}{2}+\delta}\ee
with
\be\label{8}
\rho_H=(\sqrt{\gamma^2+1}-1)\frac{R_s}{2}\ ,\quad
\delta=\frac{1}{2\sqrt{\gamma^2+1}}\ee
The remaining field equations for $B$ and $\varphi$ are given by
\be\label{9}
\frac{\partial\ln B}{\partial\ln\rho}=\frac{R_s}{A\rho}\quad,\quad
\frac{\partial\varphi}{\partial\ln\rho}=\frac{\gamma MR_s}{A\rho}
\ee
and we note that a combination of $\ln B$ and $\varphi$ is independent
of $\rho$
\be\label{10}
\ln B-\frac{\varphi}{\gamma M}=const=-\frac{\varphi_{\infty}}
{\gamma M}\ee
For large $r\gg\rho_H$ we observe the limits
\be\label{11}
\lim_{r\to\infty}\rho=r-R_s\ ,\quad \lim_{r\to\infty}B=1-\frac{R_s}{r}
+O(r^{-3})\ ,
\quad \lim_{r\to\infty}A=(1-\frac{R_s}{r})^{-1}-\frac{\gamma^2R^2_s}{4r^2}
\ee
and we identify $R_s$ with the Schwarzschild radius
\be\label{12}
R_s=2mG=\frac{m}{8\pi M^2}\ee
In lowest (nontrivial)
order post-Newtonian gravity there is no deviation from
the usual results of general relativity. In this respect the cosmon
is quite different from the Jordan-Brans-Dicke theory \cite{Wet0}, \cite{1}.
Furthermore, for $\gamma\to 0$
one has $\delta\to 1/2,\ \rho_H\to0,\ r=\rho+R_s$
and our solution approaches the well-known Schwarzschild solution
in empty space for all values of $r$.

The constant $\gamma$ reflects the role of the scalar gradient
energy density
\be\label{13}
\rho_{grad}=\frac{{\varphi'}^2}{2A}=\frac{M^2w^2}{2r^2A}=
\frac{\gamma^2R_s^2M^2}{2r^2\rho^2A}\ee
which decays for large $r$ as $\rho_{grad}\sim r^{-4}$. We
emphasize that the total mass of the object (as given by the integration
constant $R_s$) is not directly related to the scalar gradient
energy density. Indeed, gravitational and scalar energy density
are of comparable strength and may partially cancel. This effect
allows two independent integration constants even for large $\gamma$.
We also note that in linear approximation the Newtonian potential $\sim g_{
00}$ couples to $S_{00}=T_{00}-\frac{1}{2}T^\rho_\rho g_{00}$
which vanishes for a free scalar field. In absence of a potential
there is a simple mapping between solutions with $\gamma>0$ ($\varphi$
decreasing for $r\to0$) and $\gamma<0$ ($\varphi$ increasing for
$r\to 0$).

For $\gamma\not=0$ the scalar field generates a new characteristic
length scale $\rho_H$. For $r$ in the vicinity of $\rho_H+R_s$ or smaller
the overall behavior of our solution changes drastically as compared to
the Schwarzschild solution. In particular, for $r\ll \rho_H+R_s$ the
approximate behavior $(2/(1-2\delta)=2+R_s/\rho_H)$
\be\label{14}
\rho=\rho_H+(2\rho_H+R_s)^{-(1+\frac{R_s}{\rho_H})}r^{2+\frac{R_s}{\rho_H}}\ee
implies constant $\rho(r)\to\rho_H$ for $r\to0$.
As a consequence, the Schwarzschild singularity at $r=R_s$ has now
disappeared. The only possible singular behavior can occur for
$r\to0$ where
\bear\label{15}
&&\lim_{r\to0}A=\left(2+\frac{R_s}{\rho_H}\right)^{-\frac{R_s}{\rho_H}}
\left(\frac{r}{\rho_H}\right)^{2+\frac{R_s}{\rho_H}}\nonumber\\
&&\lim_{r\to 0}B\sim\left(\frac{r}{\rho_H}\right)^{\frac{R_s}{\rho_H}}\ear
Even for arbitrarily small $|\gamma|\not=0$ the black hole solution gets
modified!

An interesting situation arises for $\rho_H\gg R_s$ which corresponds
to $\gamma^2\gg1$. In this case we find a ``halo region'' with constant
rotation curve for $r\ll\rho_H$. Indeed, in this region B evolves logarithmically
\be\label{16}
B(r)\approx 1+\frac{2}{|\gamma|}\ln\frac{r}{\rho_H}\ee
The velocity of objects in stable circular orbits at distance $R$
from the center becomes therefore essentially constant
\be\label{17}
v^2_{rot}(R)=\frac{R}{2}\frac{\partial B(R)}{\partial R}\approx\frac{1}{|\gamma|}
\ee
ressembling very much the flat rotation curves observed \cite{10}
in the galactic halos!
Relating the radius of the halo to the total mass and the rotation
velocity
\be\label{18}
\rho_H\approx\frac{|\gamma|R_s}{2}\approx\frac{R_s}{2v^2_{rot}}=\frac{mG}
{v^2_{rot}}=\frac{m}{16\pi M^2v^2_{rot}}\ee
we find for a lump with $m=3\cdot 10^{11} M_{\odot}$ and $v_{rot}=150$
km/sec a halo size of around 60 kpc, well compatible\footnote{We recall
here that the mass estimates from cold dark matter models do not apply
in our case since the flattening of the rotation curve is not
due to pressureless matter. Otherwise, the currently estimated value of
$v_{rot}\approx 330$ km/sec for $m=3\cdot 10^{11}M_\odot$ would lead
to a too small halo $\sim$ 12 kpc.}
with observed
rotation curves of galaxies! These simple properties lead to the
speculation that galaxies may be cosmon lumps, with the flat rotation
curve arising through the interplay of gravitational
and scalar field equations rather than from the usually assumed dark
matter in the halo! Furthermore, for cosmon lumps generated in early cosmology
one may assume an almost constant average density of mass/halo,
typically characterized by a small power $\zeta,
m\rho_H^{-3}\sim m^\zeta$. This implies a scaling relation between
$|\gamma|$ and $m$, i.e.
\be\label{19a}
m\sim v_{rot}^{\frac{6}{2+\zeta}}\ee
and can fit the observation.

Before going on with the speculation that (some)
galaxies are cosmon lumps, one should check if
the unusual features of cosmon lumps are not in contradiction
with observation. The most striking effect is perhaps the substantial
deviation of $A$ from one inside the halo\footnote{Note that inside
the halo ${\varphi'}^2\sim r^{-2}$ and therefore $\rho_{grad}\sim r^{-4}$
(eq. (\ref{13})).}
\be\label{19}
A\approx\frac{r^2}{\rho_H^2}\ee
For local gravity  measurements at some place $R$ in the halo not too
close to the singularity the resulting effects seem to be small. We
can use a ``cartesian'' coordinate system $ds^2=-dt^2+C(\tilde\rho)d\vec x
d\vec x$ with
\be\label{20}
C(\tilde\rho)=\frac{r^2}{\tilde\rho^2}\approx\frac{\rho_H^2}{\tilde\rho^2}\  
\ln^2\left(\frac{\tilde\rho}{\tilde\rho_c}\right)\ ,\quad
\tilde\rho=\sqrt{\vec x^2}\approx\tilde\rho_c\exp\left(\frac{r}{\rho_H}\right)\ee
(where we have taken the limit $1/|\gamma|=0,\ B=1$). Choosing an appropriate
scale for $\tilde\rho$ with $C(\tilde\rho(R))=1$ the deviations from the
``local'' Minkowski metric are of the order $(\partial C/\partial\tilde
\rho)\Delta =O(\Delta\rho_H/(\tilde\rho(R)r(R)))=O(\Delta\rho_H/R^2)$ where  
$\Delta$ is
the typical length scale of the local observation. This
should be compared to the local effect of a star on the
space-like metric $\sim 2GM_\odot/\Delta$ or to the influence
of the gravitational field of the galaxy in the standard picture
$\sim O(\Delta R_s/R^2)$. As long as $\Delta^2/R^2\ll M_\odot/(\gamma m)$,
the effect is small and
local distortions of the metric by local objects can be
treated as usual. Measurements of the deflection of light by
a star or the precession of perihelia of planets would give the standard
results. The deflection of light by a distant galaxy (relevant for
gravitational lensing) is reduced
only if the light passes through the halo. Effects of this type
(also for clusters) may perhaps be used for future tests of our
speculation. The observation of rotation curves of distant galaxies
measures  the Doppler shift in frequencies $\sim R\dot\varphi=v_{rot}$
vs. the observation angle\footnote{We assume an observer in flat space.
If the observer is within a cosmon lump, the deflection of light by
its own ``galaxy'' is reduced as compared to a ``dark matter galaxy''.}
$\psi=R_{min}/d$, where $d$ is the distance to the overserver and
$R_{min}$ the value of the standard coordinate $r$ reached at the
closest distance of the photon trajectory to the center of the object.
Therefore the ``measured radius of the orbit'' corresponds to the
standard coordinate $r$,  and the function $A(r)$ plays no role.

The singularity in the center of the galaxy differs
from the standard black hole solution. This may have
interesting consequences in view of the difficulties
of standard cold dark matter galaxies to describe
the inner region. Unless $|\gamma|$ is tiny,  the scalar
field would be an important ingredient for the galactic dynamics.
This would influence many aspects of our present understanding.
The two last issues require, however, the inclusion of the baryonic
(and dark) matter
and also possibly of the scalar potential energy.

We next discuss the effects of the scalar potential. They depend
strongly on the sign of $\gamma$. We assume that at very large distances
the scalar field approaches its cosmological value. If quintessence
is important today, this implies
$V(\varphi_\infty)\approx\rho_c=6M^2H^2$, with $H$ the Hubble parameter. In
the region of small $V$ discussed previously the scalar field depends
on $r$ in a simple form
\be\label{21}
\varphi(r)\approx\left\{\begin{array}{ll}
\varphi_\infty-\frac{2M\epsilon\rho_H}{r}& {\rm outside\ the\ halo}\\
\varphi_\infty+2M\epsilon\ln\frac{r}{\rho_H}& {\rm inside\ the\ halo}
\end{array}\right.\ee
where $\epsilon=sign(\gamma)$. For $\gamma<0$ the scalar field increases for
$r\to0$. The potential energy therefore decreases and remains completely
insignificant for our discussion such that the previous discussion applies
without modifications. On the other hand, for $\gamma>0$ the potential
energy increases for decrasing $r$ and may finally become
important in the inner region of the lump. A criterion for the
importance of the potential is the relative size of the dimensionless
quantities
\be\label{22}
v=\frac{Ar^2V}{M^2}\ ,\quad y=\frac{\partial\ln B}{\partial\ln r}\ee
At large distance gravity becomes very weak (cf. eq. (11)) and $v$
dominates for $r>r_{eq}$,
\be\label{23}
r_{eq}=(m/8\pi\rho_c)^{1/3}\ee
Beyond $r_{eq}$ the metric and $\varphi$ are given by cosmology (or
the environment at larger scales like clusters).
For $m=(10^{11}-10^{12})M_\odot$ one finds that $r_{eq}=(320-700) kpc$
is outside the range of interest of this work\footnote{The
relation $\rho_c=m/(8\pi r^3_{eq})$ associates $r_{eq}$ with the
average distance between galaxies up to a factor of order one.}
(note $r_{eq}>\rho_H$). Outside the halo the change of
$V$ and $A$ are small, $v\sim r^2$, whereas inside the halo
region one has
\be\label{24}
v=\frac{r^4\rho_c}{\rho_H^2M^2}\exp(-\alpha\frac{\varphi(r)-\varphi_\infty}
{M}\Big)=6H^2\rho^2_H\left(\frac{r}{\rho_H}\right)^{4-2\epsilon\alpha}
\ee

For $\gamma>0(\epsilon=1)$, the potential becomes first negligible
for $r<r_{eq}$ until
$v$ rises again due to the decrease of $\varphi$. Then
$v$ becomes comparable to $y\approx R_s/\rho_H$ at
some radius $r_V$, provided $\alpha>2$. The scale $r_V$ depends on
$\alpha$
\be\label{25}
\frac{r_V}{\rho_H}=X_V^{\frac{1}{2\alpha-4}}\ ,\quad X_V=\frac{6H^2\rho_
H^3}{R_s}\approx 2\cdot 10^{-3}\ee
where the last number is a guideline for $m=3\cdot 10^{11}M_{\odot},
v_{rot}=150$ km/sec. For $\alpha \stackrel{\scriptstyle<}{\sim}3.5$
the ratio $r_V/\rho_H$ is smaller than 0.1. Numerically we find flat
rotation curves with $v_{rot}\leq150$ km/sec for $\alpha$ below or in the
vicinity of two. For larger $\alpha$  the rotation velocities increase
as $r$ decreases. For $\alpha\stackrel{\scriptstyle>}{\sim}2.3$ only
the solutions with $\gamma<0$ show an interesting halo
in this velocity range. For the exponential
potential (\ref{2}) we have found no solution for the gravity-scalar system
where the rotation curve decreases in some ``inner region''.

For a real galaxy baryonic matter and possibly some form of incoherent
dark matter are obviously a crucial ingredient for the dynamics of
the inner region. In this context, the incoherent dark matter
could be composed of small-size cosmon lumps
or consist of so far undiscovered particles. Indeed,
cosmon lumps can exist with various sizes and the dynamics
of small lumps may be close to a liquid of nonrelativistic particles. Dark
matter could also be mimicked by more general small wavelength
fluctuations of the cosmon field. Without
an understanding of the dynamics of matter in the inner region,
it seems premature to decide if the cosmon lump has something to do with a
galaxy. The solution presented here could be a valid description
of the halo region if galaxies are characterized by large $|\gamma|$.
In this case new particles  are not necessarily
needed for dark matter. A perhaps more conservative scenario would assume
that some form of dark matter plays an important role in the halo.
Then $|\gamma|$ could be of the order one and smaller. In this event
the interplay
between the cosmon, gravity and matter may lead to new solutions with an
effective halo region dominated by incoherent dark matter. An interesting
observational test of the ``pure cosmon halo'' discussed in
this note would compare the mass determinations of galaxies from
rotation curves  with independent ones from gravitational
lensing. For the ``cosmon halo'' the two estimates need not to coincide.

In summary, we have discussed the  isotropic solutions for a
scalar field coupled to gravity
which are relevant for extended objects in the framework of
cosmological scenarios with quintessence. We have not yet investigated  the stability
properties and the detailed character of the singularity at the origin.
Both questions depend on the so far omitted matter in the
bulk region. For a substantial cosmon density (large $|\gamma|$) one
finds a region with a flat rotation curve. It can fit with the typical
scales for mass, size and rotational
velocities of a galactic halo. Cosmon lumps should therefore
be considered as interesting candidates for
galaxies. Again, the detailed form of the rotation
curve will be influenced by the matter in the bulk region.
More generally, if quintessence
plays a role in cosmology, the effects of the cosmon may be relevant
for our understanding of extended objects as well!

\end{document}